\documentclass[12pt,letterpaper]{article}
\usepackage{epsfig,setspace,amsmath,epsf,amssymb,bm,theorem,cite, graphicx, epstopdf, algorithm, algpseudocode,float,color}
\usepackage{multirow}
\usepackage[table,xcdraw]{xcolor}

\newtheorem{theorem}{Theorem}
\newtheorem{corollary}{Corollary}

\newtheorem{remark}{Remark}
\newtheorem{lemma}{Lemma}

\newenvironment{Proof}[1]{\medskip\par\noindent{\bf Proof:\,}\,#1}{{\mbox{\,$\blacksquare$}\par}}

\allowdisplaybreaks

\begin{document}
	
\title{The Capacity of Private Information Retrieval with \\ Private Side Information Under Storage Constraints\thanks{This work was supported by NSF Grants CNS 13-14733, CCF 14-22111, CNS 15-26608 and CCF 17-13977.}}
	
\author{Yi-Peng Wei  \qquad Sennur Ulukus\\
	\normalsize Department of Electrical and Computer Engineering\\
	\normalsize University of Maryland, College Park, MD 20742 \\
	\normalsize {\it ypwei@umd.edu  \qquad \it ulukus@umd.edu} }
	
\maketitle

\begin{abstract}
We consider the problem of private information retrieval (PIR) of a single message out of $K$ messages from $N$ replicated and non-colluding databases where a cache-enabled user (retriever) of cache-size $S$ possesses side information in the form of uncoded portions of the messages that are unknown to the databases. The identities of these side information messages need to be kept private from the databases, i.e., we consider PIR with private side information (PSI). We characterize the optimal normalized download cost for this PIR-PSI problem under the storage constraint $S$ as $D^*=1+\frac{1}{N}+\frac{1}{N^2}+\dots+\frac{1}{N^{K-1-M}}+
\frac{1-r_M}{N^{K-M}}+\frac{1-r_{M-1}}{N^{K-M+1}}+\dots+\frac{1-r_1}{N^{K-1}}$, where $r_i$ is the portion of the $i$th side information message that is cached with $\sum_{i=1}^M r_i=S$. Based on this capacity result, we prove two facts: First, for a fixed memory size $S$ and a fixed number of accessible messages $M$, uniform caching achieves the lowest normalized download cost, i.e., $r_i=\frac{S}{M}$, for $i=1,\dots, M$, is optimum. Second, for a fixed memory size $S$, among all possible $K-\left \lceil{S} \right \rceil+1$ uniform caching schemes, the uniform caching scheme which caches $M=K$ messages achieves the lowest normalized download cost.
\end{abstract}

\section{Introduction}

We consider the private information retrieval (PIR) problem with private side information (PSI) for a cache-enabled user (retriever) under a cache storage size constraint. PIR refers to the problem where a user wishes to download a desired message from distributed replicated databases while keeping the identity of the desired message private against the databases. PSI refers to the setting where the user (retriever) possesses cached messages in its local storage, which it wants to utilize to decrease the download cost during PIR, but at the same time, keep their identities private against the databases. The goal of the PIR-PSI problem is to devise the most efficient retrieval scheme under the joint desired message and side information privacy constraints. The efficiency of a PIR-PSI scheme is measured by the normalized download cost which is the ratio of the number of total downloaded bits to the number of desired bits. In this work, we consider the PIR-PSI problem under a storage size constraint at the user, and investigate how best the fixed-size user cache can be utilized.

\begin{figure}[t]
	\centering
	\epsfig{file=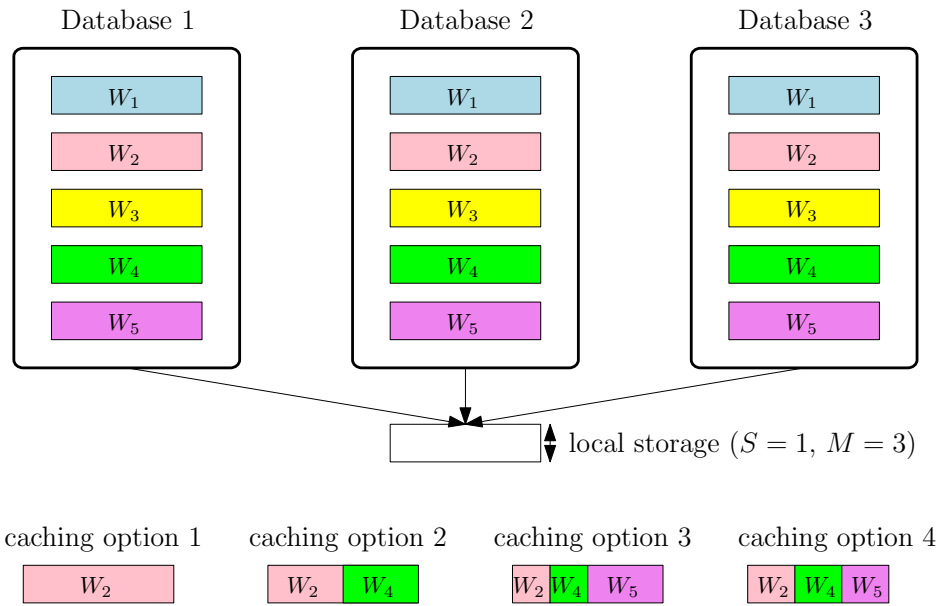,width=0.75\textwidth}
	\caption{PIR-PSI under a storage constraint. Here $N=3$, $K=5$, $S=1$, and $M=3$.}
	\label{ex_intro}
\end{figure}

We introduce the PIR-PSI problem under a storage constraint using the example shown in Fig.~\ref{ex_intro}. Consider a user wanting to download a message from $N=3$ non-communicating databases, each storing the same set of $K=5$ messages. Assume that the user is already in possession of $M=3$ messages through some unspecified means; the user may have obtained these from another user, or it may have prefetched them from another database. The databases do not know the identities of these messages, but they know that the user has access to $M=3$ messages. (For this example, say these messages are $W_2$, $W_4$ and $W_5$.) However, the user has limited local storage with size $S=1$ message. What should the user keep in order to minimize the download cost of the desired message during the PIR phase while keeping the identities of both desired and cached messages private? Should the user keep $1$ full message in its cache and discard the other 2 messages, shown as caching option $1$ in Fig.~\ref{ex_intro}? Should the user choose $2$ messages, store half of each chosen message and discard the remaining 1 message, shown as caching option $2$ in Fig.~\ref{ex_intro}? Or, should the user keep all 3 messages and store a portion of each? In that case, what portions of messages should the user store? E.g., should it store $25\%$ of $W_2$, $25\%$ of $W_4$ and $50\%$ of $W_5$, shown as caching option $3$, or should it store $\frac{1}{3}$ of all 3 messages, shown as caching option $4$ in Fig.~\ref{ex_intro}?

Different caching schemes result in different download costs for the PIR-PSI problem. Intuition may say that if portions of many messages are kept in the cache, then the user will need to protect many identities from the databases due to the PSI requirement, which may seem disadvantageous. On the other hand, intuition may also say that keeping portions of many messages may improve the diversity of side information for the PIR phase, which may seem advantageous. What is the optimum way to utilize the user's limited cache memory? In this work, we characterize the optimal normalized download cost for any given caching strategy, and determine the optimal caching strategy under a given storage constraint.

\textit{Related Work:} The PIR problem has originated in the computer science community \cite{ChorPIR, PIRsurvey2004, cachin1999computationally, ostrovsky2007survey, yekhanin2010private} and has drawn attention in the information theory society \cite{RamchandranPIR, unsynchonizedPIR, YamamotoPIR, VardyConf2015, RazanPIR, JafarConf2016} in recent years. In the classical setting of PIR, there are $N$ non-communicating databases, each storing the same set of $K$ messages. The user wishes to download one of these $K$ messages without letting the databases know the identity of the desired message. Sun and Jafar \cite{JafarPIR} have characterized the optimal normalized download cost for the classical PIR problem to be $\frac{D}{L}=\left( 1+\frac{1}{N}+\dots+\frac{1}{N^{K-1}}\right)$, where $L$ is the message size and $D$ is the total number of downloaded bits from the $N$ databases. After \cite{JafarPIR}, many interesting variants of the classical PIR problem have been investigated in \cite{JafarColluding, symmetricPIR, KarimCoded, arbmsgPIR, codedsymmetric, MultiroundPIR, codedcolluded, codedcolludedJafar, arbitraryCollusion, MPIRjournal, codedcolludingZhang, MPIRcodedcolludingZhang, BPIRjournal, tandon2017capacity, symmetricByzantine, tajeddine2017robust, wang2017linear, kadhe2017private, wei2017fundamental, chen2017capacity, wei2017capacity, wang2017secure, sun2017_computation, Kimcache2017, mirmohseni2017private, abdul2017private, wei2017jsac, banawan2018asymmetry, chen2018asymptotic, banawan2018private, d2018lifting, tajeddine2018private, wang2018secure}. The most closely related branch of PIR to our setting in this paper is cache-aided PIR in \cite{tandon2017capacity, kadhe2017private, wei2017fundamental, chen2017capacity, wei2017capacity, Kimcache2017, wei2017jsac}.


Cache-aided PIR is first considered in \cite{tandon2017capacity}, where the user has a local cache of storage $S$ messages ($SL$ bits) which can store any function of the $K$ messages, and the cache content of the user is perfectly known to all the $N$ databases. The optimal normalized download cost for this case is $D^*(S)= (1-\frac{S}{K})\left(1+\frac{1}{N}+\cdots+\frac{1}{N^{K-1}}\right)$, which indicates that the user should download the uncached part of the content via the optimum PIR scheme in \cite{JafarPIR}. The result is somewhat pessimistic since the user cannot further reduce the download cost by using the cache content. This has motivated subsequent works which have considered the case where the databases are completely unaware or partially unaware of the cache content \cite{kadhe2017private, wei2017fundamental, chen2017capacity, wei2017capacity, Kimcache2017, wei2017jsac}. Within this sub-branch of literature, references \cite{kadhe2017private,chen2017capacity,wei2017capacity} have considered PIR with PSI.

In \cite{kadhe2017private}, the authors considered the case where the user randomly chooses $M$ full messages out of $K$ messages to cache, and none of the databases is aware of the identities of the $M$ chosen messages. The user wishes to keep the identities of the $M$ chosen messages and the desired message private, which is coined as PIR with PSI. For the case of a single database, the optimal normalized download cost is settled in \cite{kadhe2017private}. For general number of databases, the optimal normalized download cost is characterized in \cite{chen2017capacity} as $D^*(M)= \left(1+\frac{1}{N}+\cdots+\frac{1}{N^{K-1-M}}\right)$. In \cite{wei2017capacity}, a more practical scenario is considered where each database is aware of the identities of the messages cached from that database only and unaware of the remaining identities of messages cached from other databases, which is coined as PIR with partially known PSI. Interestingly, the optimal normalized download cost for PIR with partially known PSI is the same as the optimal normalized download cost for PIR with PSI.

\textit{Coming back to our paper,} in this work, we consider PIR-PSI under a storage constraint. In the prefetching phase, the user can access $M$ messages, and has a local cache storage of $S$ messages ($SL$ symbols), where $S \leq M$. For each of these $M$ messages, the user caches the first $Lr_i$ symbols out of the total $L$ symbols for $i=1, \dots, M$. The caching scheme is subject to a memory size constraint, i.e., $\sum_{i=1}^M r_i=S$. Note that in \cite{wei2017fundamental, Kimcache2017, wei2017jsac}, for each message, the user randomly chooses $Lr$ symbols out of the total $L$ symbols to cache. In \cite{wei2017fundamental, Kimcache2017, wei2017jsac}, to reliably reconstruct the desired message, the user should record the indices of the cached symbols within each message. In contrast, here, we consider the case where the user caches the first $Lr_i$ symbols of each message instead of random $Lr_i$ symbols; this saves the user extra storage overhead. The databases are aware of the caching scheme, but do not know the identities of the cached messages, i.e., the databases know $M$ and $r_i$ for $i=1, \dots, M$, but do not know the identities of the cached messages. In the retrieval phase, the user wishes to jointly keep the identities of the cached messages and the desired message private. We call this model as PIR-PSI under a storage constraint.

For any given caching scheme, i.e., for given $M$ and $(r_1, r_2, \dots, r_M)$, we characterize the optimal normalized download cost to be $D^*=1+\frac{1}{N}+\frac{1}{N^2}+\dots+\frac{1}{N^{K-1-M}}
+\frac{1-r_M}{N^{K-M}}+\frac{1-r_{M-1}}{N^{K-M+1}}+\dots+\frac{1-r_1}{N^{K-1}}$, where without loss of generality $r_1 \geq r_2 \geq \dots \geq r_M$. Based on this capacity result, we prove two important facts: First, for a fixed memory size $S$ and fixed number of accessible messages $M$, uniform caching achieves the lowest normalized download cost, where uniform caching means $r_i=\frac{S}{M}$, for $i=1,\dots, M$. Second, for a fixed memory size $S$, among all the $K-\left \lceil{S} \right \rceil+1$ uniform caching schemes, the uniform caching scheme which caches $K$ messages achieves the lowest normalized download cost. That is, in order to optimally utilize the limited user cache memory, if the user has access to $M$ files, it should keep $SL/M$ bits (equal amounts) from each message in its cache memory; and second, if possible, the user should aim to have access to all $K$ messages, i.e., $M=K$ yields the lowest download cost.

\section{System Model}

We consider a system consisting of $N$ non-communicating databases and a user (retriever). Each database stores the same set of $K$ independent messages $W_1, \dots, W_K$, and each message is of size $L$ symbols, i.e.,
\begin{align}
H(W_1)=\dots=H(W_K)=L, \qquad H(W_1, \dots, W_K)=H(W_1)+\dots+H(W_K).
\end{align}
The user has a local cache memory which is of size $SL$ symbols, where $ S \in [0, K]$. There are two phases in the system: the prefetching phase and the retrieval phase.

In the prefetching phase, the user can randomly access $M$ messages out of total $K$ messages, where $M \geq S$. For each of the $M$ accessed messages, the user caches the first $Lr_i$ symbols out of the total $L$ symbols for $i=1, \dots, M$. The caching scheme is subject to a memory size constraint of $S$, i.e.,
\begin{align}
\sum_{i=1}^M r_i=S.
\end{align}
We denote the indices (identities) of the cached $M$ messages as $\mathbb{H}$, and denote $\mathcal{W}_\mathbb{H}$ as the cached messages. Therefore, $|\mathbb{H}|=M$, and $H(\mathcal{W}_\mathbb{H})=SL$.

Note that $M$ and $(r_1, \dots, r_M)$ specify a caching scheme. If $r_1=\dots=r_M$, we call this a uniform caching scheme. For fixed $S$, there are $K-\left \lceil{S} \right \rceil+1$ uniform caching schemes depending on the number of accessible messages since $M\geq S$. For instance, if there are $K=3$ messages in the databases and $S=1.5$, then since $M\geq S$, $M$ can take one of two possible values: either 2 or 3. Thus, there are two uniform caching schemes depending on the value of $M$. Note, $K-\left \lceil{S} \right \rceil+1=3-\left \lceil 1.5 \right \rceil +1=2$.

We assume that all the databases are aware of the caching scheme but are unaware of which messages are cached. For example, if $S=2$, $M=3$, and we say that the user has applied a uniform caching scheme, the databases know that the user has chosen $3$ messages out of the total $K$ messages to cache, and for each chosen message, the user has cached the first $\frac{2}{3}L$ symbols out of the total $L$ symbols. However, the databases do not know which messages are cached by the user.

In the retrieval phase, the user privately generates an index $\theta \in [K]=\{1, \dots, K\}$, and wishes to retrieve message $W_\theta$ such that it is impossible for any individual database to identify $\theta$. At the same time, the user also wishes to keep the indices of the $M$ cached messages private, i.e., in the retrieval phase the databases cannot learn which messages are cached. For random variables $\theta$, $\mathbb{H}$, and $W_1,\dots,W_K$, we have
\begin{align} \label{independency}
H\left(\theta, \mathbb{H}, W_1,\dots,W_K  \right)= H\left( \theta \right) + H\left( \mathbb{H} \right) + H(W_1)+\dots+H(W_K).
\end{align}

In order to retrieve message $W_\theta$, the user sends $N$ queries $Q_1^{[\theta, \mathbb{H}]}, \dots, Q_N^{[\theta, \mathbb{H}]}$ to the $N$ databases, where $Q_n^{[\theta, \mathbb{H}]}$ is the query sent to the $n$th database for message $W_\theta$. Note that the queries are generated according to $\mathbb{H}$, which are independent of the realization of the $K$ messages. Therefore,
\begin{align} \label{query_indep}
I(W_1, \dots, W_K;  Q_1^{[\theta, \mathbb{H}]}, \dots,  Q_N^{[\theta,\mathbb{H} ]}  ) =0.
\end{align}
Upon receiving the query $Q_n^{[\theta, \mathbb{H}]}$, the $n$th database replies with an answering string $A_n^{[\theta, \mathbb{H}]}$, which is a function of  $Q_n^{[\theta, \mathbb{H}]}$ and all the $K$ messages. Therefore, $\forall \theta \in [K], \forall n \in [N]$,
\begin{align} \label{answer_constraint}
H(A_n^{[\theta, \mathbb{H}]}|Q_n^{[\theta, \mathbb{H}]}, W_1, \dots, W_K)=0.
\end{align}
After receiving the answering strings $A_1^{[\theta, \mathbb{H}]}, \dots, A_N^{[\theta, \mathbb{H}]}$ from all the $N$ databases, the user needs to decode the desired message $W_\theta$ reliably. By using Fano's inequality, we have the following reliability constraint
\begin{align} \label{reliability_constraint}
H\left(W_\theta|\mathcal{W}_\mathbb{H}, \mathbb{H}, Q_1^{[\theta,\mathbb{H}]}, \dots, Q_N^{[\theta,\mathbb{H}]}, A_1^{[\theta,\mathbb{H}]}, \dots, A_N^{[\theta,\mathbb{H}]} \right) = o(L),
\end{align}
where $o(L)$ denotes a function such that $\frac{o(L)}{L} \rightarrow 0$ as $L \rightarrow \infty$.

To ensure that individual databases do not know which message is retrieved and to keep the $M$ cached messages private, we have the following privacy constraint, $\forall n \in [N]$, $\forall \theta, \theta' \in [K]$, $\forall \mathbb{H}, \mathbb{H}' \subset [K]$ such that $|\mathbb{H}|=|\mathbb{H}'|=M$,
\begin{align} \label{privacy_constraint}
(Q_n^{[\theta,\mathbb{H}]}, A_n^{[\theta,\mathbb{H}]}, W_1, \dots, W_K)  \sim (Q_n^{[\theta',\mathbb{H}']}, A_n^{[\theta',\mathbb{H}']}, W_1, \dots, W_K) ,
\end{align}
where $A \sim B$ means that $A$ and $B$ are identically distributed.

For a fixed $N$, $K$, $S$ and caching scheme $(r_1, \dots, r_M)$, a pair $\left(D,L\right)$ is achievable if there exists a PIR scheme for the message which is of size $L$ symbols satisfying the reliability constraint \eqref{reliability_constraint} and the privacy constraint \eqref{privacy_constraint}, where $D$ represents the expected number of downloaded bits (over all the queries) from the $N$ databases via the answering strings $A_{1:N}^{[\theta,\mathbb{H}]}$, where $A_{1:N}^{[\theta,\mathbb{H}]}=(A_1^{[\theta,\mathbb{H}]}, \dots, A_N^{[\theta,\mathbb{H}]} )$, i.e.,
\begin{align}
D=\sum_{n=1}^N H\left(A_n^{[\theta,\mathbb{H} ]} \right).
\end{align}
In this work, we aim at characterizing the optimal normalized download cost $D^*$, where
\begin{align}
D^*=\inf \left\{ \frac{D}{L}: \left(D, L \right) \text{ is achievable}  \right\}.
\end{align}

\section{Main Results and Discussions}

We characterize the exact normalized download cost for PIR-PSI under a storage constraint in the following theorem.

\begin{theorem} \label{Thm1}
In PIR-PSI under a storage constraint, the optimal normalized download cost is
\begin{align}
D^*=1+\frac{1}{N}+\frac{1}{N^2}+\dots+\frac{1}{N^{K-1-M}}+\frac{1-r_M}{N^{K-M}}+\frac{1-r_{M-1}}{N^{K-M+1}}
+\dots+\frac{1-r_1}{N^{K-1}} \label{eq_main2}
\end{align}
where $r_1 \geq r_2 \geq \dots \geq r_M$ without loss of generality.
\end{theorem}

The converse proof for Theorem~\ref{Thm1} is given in Section~\ref{Sec_Conv}, and the achievability proof for Theorem~\ref{Thm1} is given in Section~\ref{Sec_Achv}.

\begin{remark}
For $S=0$, by letting $r_i=0$, for $i=1, \dots, M$, \eqref{eq_main2} reduces to
\begin{align} \label{eq_PIR}
D^*= 1+\frac{1}{N}+\frac{1}{N^2}+\dots+\frac{1}{N^{K-1}},
\end{align}
which is the optimal normalized download cost of the original PIR problem as shown in \cite{JafarPIR}.
\end{remark}

\begin{remark}
For $S\in[K]$ and $M=S$, by letting $r_i=1$ for $i=1,\dots, M$, \eqref{eq_main2} reduces to
\begin{align} \label{eq_MDS_PIR}
D^*= 1+\frac{1}{N}+\frac{1}{N^2}+\dots+\frac{1}{N^{K-1-M}},
\end{align}
which is the optimal normalized download cost of the PIR with PSI problem as shown in \cite{chen2017capacity}. We can further generalize the result to the PIR with partially known PSI as shown in \cite{wei2017capacity}. Note further that for $M>S$, $(\frac{1-r_M}{N^{K-M}}+\frac{1-r_{M-1}}{N^{K-M+1}}+\dots+\frac{1-r_1}{N^{K-1}})$ is the penalty to the download cost under the storage constraint.
\end{remark}

\begin{corollary} \label{cor_1}
For fixed $M \geq S$, uniform caching scheme achieves the lowest normalized download cost.
\end{corollary}

\begin{Proof}
The user has access to $M$ messages. To achieve a low normalized download cost in \eqref{eq_main2}, we need to solve the following optimization problem,
\begin{align}
\min_{\alpha_i, i=1, \dots, M}& \quad \alpha_M \frac{1}{N^{K-M}} + \alpha_{M-1} \frac{1}{N^{K-M+1}} + \dots + \alpha_1 \frac{1}{N^{K-1}}  \notag \\
\text{s.t.} \quad & \quad  \alpha_M+\alpha_{M-1}+\dots + \alpha_1=M-S, \notag \\
           & \quad  1\geq \alpha_M\geq \alpha_{M-1} \geq \dots \geq \alpha_1 \geq 0,
           \label{opt-prob}   						
\end{align}
which is obtained by replacing $1-r_i$ in \eqref{eq_main2} with $\alpha_i$ for $i=1,\dots, M$. We prove by contradiction that the minimum is achieved when $\alpha_M=\alpha_{M-1}$. Suppose not, then we have optimum $\alpha^*_M>\alpha^*_{M-1}$. Choose $\delta=\frac{\alpha^*_M-\alpha^*_{M-1}}{3}$, and let $\alpha'_M=\alpha^*_M-\delta$, $\alpha'_{M-1}=\alpha^*_{M-1}+\delta$. Then, with $\alpha'_M$ and $\alpha'_{M-1}$, we achieve a lower normalized download cost than with $\alpha^*_M$ and $\alpha^*_{M-1}$, which gives a contradiction. Therefore, we have $\alpha_M=\alpha_{M-1}$. Intuitively, note that the coefficient of $\alpha_{M}$ is larger than the coefficient of $\alpha_{M-1}$ in the objective function in (\ref{opt-prob}). Therefore, in order to minimize the objective function, we need to choose $\alpha_{M}$ as small as possible. But, since $\alpha_M$ needs to be larger than $\alpha_{M-1}$ according to the constraint set of (\ref{opt-prob}), the smallest $\alpha_M$ we can choose is $\alpha_M=\alpha_{M-1}$. Using similar arguments, we also have $\alpha_{M-1}=\alpha_{M-2}=\dots=\alpha_{1}$. Therefore, uniform caching achieves the lowest normalized download cost for fixed $M$.
\end{Proof}

\begin{corollary} \label{cor_2}
For fixed $S$, among all the $K-\left \lceil{S} \right \rceil+1$ uniform caching schemes, the uniform caching scheme with $M=K$ achieves the lowest normalized download cost.
\end{corollary}

\begin{Proof}
For the uniform caching scheme $M$, the user caches the first $\frac{S}{M}L$ symbols of each chosen message. From \eqref{eq_main2}, the normalized download cost is
\begin{align}
D^*(M)=1+\frac{1}{N}+\frac{1}{N^2}+\dots+\frac{1}{N^{K-1-M}}+\left(1-\frac{S}{M}\right)\left(\frac{1}{N^{K-M}} +\dots + \frac{1}{N^{K-1}} \right).
\end{align}
Considering the difference of the normalized download costs between $D^*(M+1)$ and $D^*(M)$,
\begin{align}
&D^*(M+1)-D^*(M) \notag \\
&\qquad= 1+\frac{1}{N}+\frac{1}{N^2}+\dots+\frac{1}{N^{K-2-M}}+\left(1-\frac{S}{M+1}\right)\left(\frac{1}{N^{K-M-1}} +\dots + \frac{1}{N^{K-1}} \right) \notag \\
&\qquad~~ - \left[1+\frac{1}{N}+\frac{1}{N^2}+\dots+\frac{1}{N^{K-1-M}}
+\left(1-\frac{S}{M}\right)\left( \frac{1}{N^{K-M}} +\dots + \frac{1}{N^{K-1}}     \right)     \right]   \\
&\qquad= -\frac{S}{M+1} \left(\frac{1}{N^{K-M-1}} +\dots + \frac{1}{N^{K-1}} \right)
         +\frac{S}{M}\left( \frac{1}{N^{K-M}} +\dots + \frac{1}{N^{K-1}}     \right) \\
&\qquad= -\frac{S}{M+1}\times\frac{1}{N^{K-M-1}} +\left(\frac{S}{M}-\frac{S}{M+1}  \right)
\left( \frac{1}{N^{K-M}} +\dots + \frac{1}{N^{K-1}}     \right)        \\
&\qquad= \frac{S}{M(M+1)} \left( \frac{1}{N^{K-M}} +\dots + \frac{1}{N^{K-1}}     \right)
-\frac{S}{M(M+1)} \times\frac{M}{N^{K-M-1}} \\
&\qquad \leq 0.
\end{align}
Thus, the uniform caching scheme with $M=K$ achieves the lowest normalized download cost among all possible uniform caching schemes.
\end{Proof}

\begin{corollary}
For fixed $S$, among all possible caching schemes, the uniform caching scheme with $M=K$ achieves the lowest normalized download cost.
\end{corollary}

\begin{Proof}
From Corollary~\ref{cor_1}, we know that for fixed $M$, uniform caching scheme achieves the lowest normalized download cost. From Corollary~\ref{cor_2}, we know that among all uniform caching schemes, the uniform caching scheme with $M=K$ achieves the lowest normalized download cost. Combining these two corollaries, we conclude that among all possible caching schemes, the uniform caching scheme with $M=K$ achieves the lowest normalized download cost.
\end{Proof}

\section{Converse Proof} \label{Sec_Conv}	

In this section, we provide a lower bound for PIR-PSI under a storage constraint. In the following, without loss of generality, we relabel the messages according to $\mathbb{H}$, such that $W_{1:M}$ are the messages accessed by the user in the prefetching phase, where $W_{1:M}=(W_1, W_2, \dots, W_M)$. Here, $W_{i}$ denotes the message whose first $Lr_i$ symbols are cached by the user, for $i=1, 2, \dots, M$, and without loss of generality, $r_1 \geq r_2 \geq \dots \geq r_M$.

We first need the following lemma, which develops a lower bound on the length of the undesired portion of the answering strings as a consequence of the privacy constraint.

\begin{lemma}[Interference lower bound]\label{lemma_converse1}
For PIR-PSI under a storage constraint, the interference from undesired messages within the answering strings, $D-L$, is lower bounded by, 	
\begin{align} \label{eq_L1}
D-L+o(L)\geq I\left(W_{1:K-1};Q_{1:N}^{[K,\mathbb{H}]},A_{1:N}^{[K,\mathbb{H}]}|\mathcal{W}_{\mathbb{H}},\mathbb{H},W_K \right).
\end{align}
\end{lemma}

If the privacy constraint is absent, the user downloads only $L$ symbols of the desired message, however, when the privacy constraint is present, it should download $D$ symbols. The difference between $D$ and $L$, i.e., $D-L$, corresponds to the undesired portion of the answering strings. Note that Lemma~\ref{lemma_converse1} is an extension of \cite[Lemma~5]{JafarPIR}, where $M=0$, i.e., the user has no PSI. Lemma~\ref{lemma_converse1} differs from its counterpart in \cite[Lemma~1]{wei2017fundamental} in two aspects; first, the left hand side is $D(r)-L(1-r)$ in \cite{wei2017fundamental} as the user requests to download the uncached bits only, and second,  \cite[Lemma~1]{wei2017fundamental} constructs $K-1$ distinct lower bounds by changing $k$, in contrast to only one bound here. In addition, we note that a similar argument to Lemma~\ref{lemma_converse1} can be implied from \cite{chen2017capacity} and \cite{wei2017capacity}. The main difference between Lemma~\ref{lemma_converse1} and \cite{chen2017capacity,wei2017capacity} is that $\mathcal{W}_{\mathbb{H}}$ refers to parts of messages here, while in \cite{chen2017capacity,wei2017capacity}, $\mathcal{W}_{\mathbb{H}}$ refers to full messages.

\begin{Proof}
We start with the right hand side of \eqref{eq_L1},	
\begin{align}
I\left(W_{1:K-1};Q_{1:N}^{[K, \mathbb{H}]}, A_{1:N}^{[K, \mathbb{H}]}|\mathcal{W}_{\mathbb{H}},\mathbb{H},W_K \right)
\leq I\left(W_{1:K-1};Q_{1:N}^{[K,\mathbb{H}]},A_{1:N}^{[K,\mathbb{H}]},W_K|\mathcal{W}_{\mathbb{H}}, \mathbb{H} \right).  \label{eq_lemma_1}
\end{align}	
For the right hand side of \eqref{eq_lemma_1}, we have
\begin{align}
& I\left(W_{1:K-1};Q_{1:N}^{[K,\mathbb{H}]},A_{1:N}^{[K,\mathbb{H}]},W_K|\mathcal{W}_{\mathbb{H}}, \mathbb{H} \right) \notag \\
&\qquad =I\left(W_{1:K-1};Q_{1:N}^{[K,\mathbb{H}]}, A_{1:N}^{[K, \mathbb{H}]}|\mathcal{W}_{\mathbb{H}}, \mathbb{H} \right)
 + I \left( W_{1:K-1}; W_K|Q_{1:N}^{[K, \mathbb{H}]}, A_{1:N}^{[K, \mathbb{H}]},\mathcal{W}_\mathbb{H}, \mathbb{H} \right) \\
&\qquad\label{eq_ILB_1}\stackrel{\eqref{reliability_constraint}}{=}
I\left(W_{1:K-1};Q_{1:N}^{[K,\mathbb{H}]}, A_{1:N}^{[K, \mathbb{H}]}|\mathcal{W}_{\mathbb{H}}, \mathbb{H} \right)+ o(L) \\
&\label{eq_ILB_2} \quad~\stackrel{\eqref{independency},\eqref{query_indep}}{=}
I\left(W_{1:K-1}; A_{1:N}^{[K, \mathbb{H}]} |\mathcal{W}_{\mathbb{H}}, \mathbb{H}, Q_{1:N}^{[K,\mathbb{H}]} \right) + o(L) \\
&\qquad= H\left( A_{1:N}^{[K, \mathbb{H}]} |\mathcal{W}_{\mathbb{H}}, \mathbb{H}, Q_{1:N}^{[K,\mathbb{H}]} \right)
  - H\left(A_{1:N}^{[K, \mathbb{H}]} |\mathcal{W}_\mathbb{H}, \mathbb{H}, Q_{1:N}^{[K,\mathbb{H}]},W_{1:K-1}  \right)   + o(L) \\
&\qquad\leq D- H\left(A_{1:N}^{[K, \mathbb{H}]} |\mathcal{W}_\mathbb{H}, \mathbb{H}, Q_{1:N}^{[K,\mathbb{H}]},W_{1:K-1}  \right)  + o(L) \label{eq_L1_1} \\
&\qquad\label{eq_ILB_3}\stackrel{\eqref{reliability_constraint}}{=}
D-H\left(A_{1:N}^{[K, \mathbb{H}]}, W_K |\mathcal{W}_\mathbb{H}, \mathbb{H}, Q_{1:N}^{[K,\mathbb{H}]},W_{1:K-1}  \right)+o(L) \\
&\qquad\label{eq_ILB_4}\leq D-H\left(W_K |\mathcal{W}_\mathbb{H}, \mathbb{H}, Q_{1:N}^{[K,\mathbb{H}]},W_{1:K-1}  \right)  + o(L) \\
&\quad~\label{eq_ILB_5}\stackrel{\eqref{independency},\eqref{query_indep}}{=} D-L + o(L)
\end{align}
where \eqref{eq_ILB_1}, \eqref{eq_ILB_3} follow from the decodability of $W_{K}$ given $\left(Q_{1:N}^{[K, \mathbb{H}]}, A_{1:N}^{[K, \mathbb{H}]},\mathcal{W}_\mathbb{H},\mathbb{H} \right)$, \eqref{eq_ILB_2}, \eqref{eq_ILB_5} follow from the independence of $W_{1:K}$ and $Q_{1:N}^{[K, \mathbb{H}]}$ given  $\mathbb{H}$, and \eqref{eq_L1_1} follows from the independence bound. Combining \eqref{eq_lemma_1} and \eqref{eq_ILB_5} yields \eqref{eq_L1}.
\end{Proof}

For the conditional mutual information term on the right hand side of \eqref{eq_L1}, we have
\begin{align}
I\left(W_{1:K-1};Q_{1:N}^{[K,\mathbb{H}]},A_{1:N}^{[K,\mathbb{H}]}|\mathcal{W}_{\mathbb{H}},\mathbb{H},W_K \right) 
& = \sum_{h} p(h) I\left(W_{1:K-1};Q_{1:N}^{[K,h]},A_{1:N}^{[K,h]}|\mathcal{W}_h,h,W_K \right) \\
& = \sum_{h} p(h) I\left(W_{1:K-1};Q_{1:N}^{[K,h]},A_{1:N}^{[K,h]}|\mathcal{W}_h,W_{K} \right). \label{eq_conv_1}
\end{align}
where we have written the mutual information in \eqref{eq_L1} as an expectation over all possible caching scheme realizations, as the databases do not know which messages are cached.

In the following lemma, we develop an inductive relation for the mutual information term on the right hand side of \eqref{eq_conv_1}.

\begin{lemma}[Fractional induction lemma]\label{lemma_converse2}
For all $k\in \{1,\dots,K-1\}$, the mutual information term in \eqref{eq_conv_1} can be inductively lower bounded as,
\begin{align} \label{eq_L2}
&I\left( W_{1:k}; Q_{1:N}^{[k+1, h]}, A_{1:N}^{[k+1, h]} |\mathcal{W}_h, W_{k+1:K} \right) \notag \\
&\qquad\quad \geq \frac{1}{N}  I\left(W_{1:k-1} ;Q_{1:N}^{[k,h]}, A_{1:N}^{[k, h]}|\mathcal{W}_h, W_{k:K}  \right)  +\frac{L}{N}(1-r_k)- o(L),
\end{align}	
where $r_k=0$ when $k>M$.
\end{lemma}

Lemma~\ref{lemma_converse2} is a generalization of \cite[Lemma~6]{JafarPIR} to our setting. The main difference between Lemma~\ref{lemma_converse2} and \cite[Lemma~6]{JafarPIR} is that the cached PSI results in a different induction relation.

\begin{Proof}
We start with the left hand side of \eqref{eq_L2},		
\begin{align}
&I\left( W_{1:k}; Q_{1:N}^{[k+1, h]}, A_{1:N}^{[k+1, h]} |\mathcal{W}_h, W_{k+1:K} \right) \notag \\
&\qquad=\frac{1}{N} \times N \times I\left( W_{1:k}; Q_{1:N}^{[k+1, h]}, A_{1:N}^{[k+1, h]} |\mathcal{W}_h, W_{k+1:K} \right)  \\
&\qquad \geq \frac{1}{N}\sum_{n=1}^N I\left( W_{1:k}; Q_n^{[k+1, h]}, A_n^{[k+1, h]} |\mathcal{W}_h, W_{k+1:K} \right)  \label{eq_L2_1} \\
&\qquad \stackrel{\eqref{privacy_constraint}}{=}
\frac{1}{N} \sum_{n=1}^N I\left( W_{1:k}; Q_n^{[k, h]}, A_n^{[k, h]} |\mathcal{W}_h, W_{k+1:K} \right)   \label{eq_L2_2}\\
&\qquad \geq \frac{1}{N} \sum_{n=1}^N I\left( W_{1:k}; A_n^{[k, h]} |\mathcal{W}_h, W_{k+1:K}, Q_n^{[k, h]} \right) \label{eq_L2_3}  \\
&\qquad\stackrel{\eqref{answer_constraint}}{=}
\frac{1}{N}\sum_{n=1}^N H \left(A_n^{[k, h]} |\mathcal{W}_h, W_{k+1:K}, Q_n^{[k, h]} \right)\label{eq_L2_4}\\
&\qquad \geq \frac{1}{N}\sum_{n=1}^N  H \left(A_n^{[k, h]} |\mathcal{W}_h, W_{k+1:K}, Q_{1:N}^{[k, h]}, A_{1:n-1}^{[k, h]} \right)
\label{eq_L2_5}\\
&\qquad \stackrel{\eqref{answer_constraint}}{=}
\frac{1}{N}\sum_{n=1}^N I \left(W_{1:k}; A_n^{[k, h]} |\mathcal{W}_h, W_{k+1:K}, Q_{1:N}^{[k, h]}, A_{1:n-1}^{[k, h]} \right) \label{eq_L2_6}\\
&\qquad = \frac{1}{N} I \left(W_{1:k};  A_{1:N}^{[k, h]} |\mathcal{W}_h, W_{k+1:K}, Q_{1:N}^{[k, h]} \right) \\
&\quad~\stackrel{\eqref{independency},\eqref{query_indep}}{=}
\frac{1}{N} I \left(W_{1:k}; Q_{1:N}^{[k, h]},  A_{1:N}^{[k, h]} |\mathcal{W}_h, W_{k+1:K}  \right) \label{eq_L2_7}\\
&\qquad\stackrel{\eqref{reliability_constraint}}{=} \frac{1}{N}
I \left(W_{1:k};W_k, Q_{1:N}^{[k, h]},  A_{1:N}^{[k, h]} |\mathcal{W}_h, W_{k+1:K}  \right)  -o(L)   \label{eq_L2_8}\\
&\qquad = \frac{1}{N} I \left(W_{1:k}; W_k|\mathcal{W}_h, W_{k+1:K}  \right)
        + \frac{1}{N} I \left(W_{1:k}; Q_{1:N}^{[k, h]},  A_{1:N}^{[k, h]}|\mathcal{W}_h, W_{k:K}  \right)   -o(L) \\
&\qquad = \frac{1}{N} I \left(W_{1:k}; Q_{1:N}^{[k, h]},  A_{1:N}^{[k, h]}|\mathcal{W}_h, W_{k:K}  \right) +\frac{L}{N}(1-r_k) -o(L), \label{eq_L2_9}
\end{align}
where \eqref{eq_L2_1} and \eqref{eq_L2_3} follow from the chain rule and the non-negativity of mutual information, \eqref{eq_L2_2} follows from the privacy constraint,  \eqref{eq_L2_4}, \eqref{eq_L2_6} follow from the fact that answer strings are deterministic functions of the messages and the queries, \eqref{eq_L2_5} follows from the fact that conditioning reduces entropy, \eqref{eq_L2_7} follows from the independence of $W_{1:K}$ and $Q_{1:N}^{[k, h]}$, \eqref{eq_L2_8} follows from the reliability constraint on $W_k$, and \eqref{eq_L2_9} is due to the fact that $H\left(W_k|\mathcal{W}_h, W_{k+1:K}  \right)=L(1-r_k)$, where if $k\notin h$ then $r_k=0$.
\end{Proof}

By applying Lemma~\ref{lemma_converse2} recursively to the right hand side of \eqref{eq_conv_1}
\begin{align}
&I\left(W_{1:K-1};Q_{1:N}^{[K,h]},A_{1:N}^{[K,h]}|\mathcal{W}_h,W_{K} \right) \notag \\
&\qquad \stackrel{\eqref{eq_L2}}{\geq}
\frac{1}{N} I\left(W_{1:K-2} ;Q_{1:N}^{[K-1,h]}, A_{1:N}^{[K-1, h]}|\mathcal{W}_h, W_{K-1:K}  \right)+\frac{L}{N}- o(L) \label{eqc1} \\
&\qquad \stackrel{\eqref{eq_L2}}{\geq}
\frac{1}{N^2} I\left(W_{1:K-3} ;Q_{1:N}^{[K-2,h]}, A_{1:N}^{[K-2, h]}|\mathcal{W}_h, W_{K-2:K}  \right)+\frac{L}{N^2}+\frac{L}{N}- o(L) \\
&\qquad \stackrel{\eqref{eq_L2}}{\geq} \dots \\
&\qquad \stackrel{\eqref{eq_L2}}{\geq}
\frac{1}{N^{K-1-M}} I\left(W_{1:M} ;Q_{1:N}^{[M+1,h]}, A_{1:N}^{[M+1, h]}|\mathcal{W}_h, W_{M+1:K}  \right) \notag \\
&\qquad \qquad +\frac{L}{N^{K-1-M}} +\dots+ \frac{L}{N^2}+\frac{L}{N}- o(L) \label{eqc2} \\
&\qquad \stackrel{\eqref{eq_L2}}{\geq}
\frac{1}{N^{K-M}} I\left(W_{1:M-1} ;Q_{1:N}^{[M,h]}, A_{1:N}^{[M, h]}|\mathcal{W}_h, W_{M:K}  \right) +\frac{L}{N^{K-M}}(1-r_M) \notag \label{eqc3} \\
&\qquad \qquad +\frac{L}{N^{K-1-M}} +\dots+ \frac{L}{N^2}+\frac{L}{N}- o(L) \\
&\qquad \stackrel{\eqref{eq_L2}}{\geq} \dots \\
&\qquad \stackrel{\eqref{eq_L2}}{\geq}
\frac {L(1-r_1)}{N^{K-1}}+\dots+\frac{L(1-r_{M})}{N^{K-M}}+\dots+ \frac{L}{N^2}+\frac{L}{N}-o(L). \label{eqc4}
\end{align}
Note that in \eqref{eqc1} to \eqref{eqc2}, we apply the fractional induction lemma with $r=0$, since $W_{M+1:K}$ are not cached in $\mathcal{W}_h$. In \eqref{eqc3} to \eqref{eqc4}, $r_k>0$ for the fractional induction lemma, since $W_{1:M}$ are cached in $\mathcal{W}_h$ partially.

By combining \eqref{eq_L1}, \eqref{eq_conv_1}, and \eqref{eqc4}, and dividing by $L$ on both sides, we obtain a lower bound for the normalized download cost as
\begin{align}
D^* \geq 1+\frac{1}{N}+\frac{1}{N^2}+\dots+\frac{1}{N^{K-1-M}}+\frac{1-r_M}{N^{K-M}}+\frac{1-r_{M-1}}{N^{K-M+1}}+\dots+\frac{1-r_1}{N^{K-1}},
\end{align}
which proves \eqref{eq_main2}.

\section{Achievability Proof}\label{Sec_Achv}	
	
Our achievability scheme is based on the PIR schemes in \cite{JafarPIR} and \cite{chen2017capacity}. For the portion of the messages not cached by the user, we use the PIR scheme in \cite{JafarPIR}, which applies the following three principles recursively: 1) database symmetry, 2) message symmetry within each database, and 3) exploiting undesired messages as side information. For the portion of the messages cached by the user, we use the PIR scheme in \cite{chen2017capacity}, which is based on MDS codes and consists of two stages: The first stage determines the systematic part of the MDS code according to the queries generated in \cite{JafarPIR}. In the second stage, the user reduces the download cost by downloading the parity part of the MDS code only. By applying the two PIR schemes, the user retrieves the desired message privately while keeping the cached messages private.

\subsection{Motivating Examples}

\subsubsection{$N=2$ Databases, $K=5$ Messages, $M=2$ Accessed Messages, and $S=1$ with Uniform Caching}

In this example, in the prefetching phase, the user randomly chooses two messages to cache, say $W_1$ and $W_4$. Since $S=1$ and the user uses uniform caching scheme, the user caches the first half of $W_1$ and the first half of $W_4$. We note that the databases are aware of the caching scheme, i.e., the databases know that two out of five messages are chosen by the user, and the first halves of the chosen messages are cached. However, the databases do not know which are the two chosen messages.

In the retrieval phase, assume that the user wishes to retrieve message $W_3$ privately. For the first half portion of the message, i.e., for the symbols in the interval $[0, \frac{L}{2}]$, since the user has cached messages $W_1$ and $W_4$, the user applies the PIR scheme in \cite{chen2017capacity} with $M=2$. The total download cost for the first half portion of the message, as shown in \eqref{eq_MDS_PIR}, is
\begin{align} \label{eq_ex1}
\frac{L}{2} \times \left(1+\frac{1}{2}+\frac{1}{2^{5-1-2}} \right).
\end{align}
For the remaining half portion of the message, i.e., for the symbols in the interval $[\frac{L}{2}, L]$, since the user has not cached any messages, the user applies the PIR scheme in \cite{JafarPIR}. The total download cost for the remaining half portion of the message, as shown in \eqref{eq_PIR}, is
\begin{align} \label{eq_ex2}
\frac{L}{2} \times \left(1+\frac{1}{2}+\frac{1}{2^2}+\frac{1}{2^3}+\frac{1}{2^{5-1}} \right).
\end{align}

The overall download cost is the sum of \eqref{eq_ex1} and \eqref{eq_ex2}. Therefore, the optimal normalized download cost is $\frac{59}{32}$, which can also be obtained through \eqref{eq_main2} by letting $r_1=\frac{1}{2}$ and $r_2=\frac{1}{2}$. Note that since we have applied the PIR scheme in \cite{chen2017capacity} to retrieve the first half portion of the message, the databases cannot learn which messages are cached by the user. In addition, both PIR schemes in \cite{JafarPIR} and \cite{chen2017capacity} keep the identity of the desired message private. Therefore, the combination of these two PIR schemes is a feasible PIR scheme for PIR-PSI a under storage constraint \cite{arbmsgPIR}.

\subsubsection{$N=2$ Databases, $K=5$ Messages, $S=1$, $M=3$ with $r_1=\frac{1}{2}$, and $r_2=r_3=\frac{1}{4}$}

\begin{figure}[t]
	\centering
	\epsfig{file=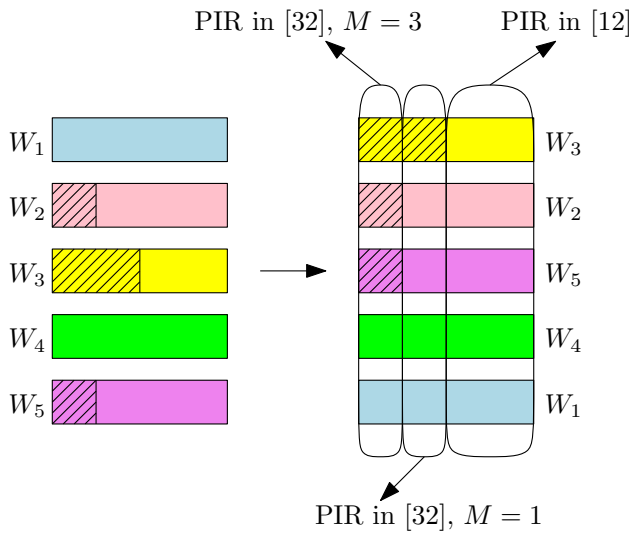,width=0.5\textwidth}
	\caption{Achievable scheme: $K=5$, $S=1$, and $M=3$ with $r_1=\frac{1}{2}$, and $r_2=r_3=\frac{1}{4}$.}
	\label{ex_ach}
\end{figure}

In this example, see Fig.~\ref{ex_ach}, in the prefetching phase, since $r_1=\frac{1}{2}$, the user first randomly chooses one message to cache, say $W_3$, and the user caches the first half of $W_3$. Since $r_2=r_3=\frac{1}{4}$, the user then randomly chooses two other messages to cache, say $W_2$ and $W_5$, and the user caches the first $\frac{1}{4}$ portions of $W_2$ and $W_5$. Note that $S=1$ and $\frac{1}{2}\times 1 + \frac{1}{4}\times 2=1$, and the local cache memory size constraint is satisfied. We note that the databases are aware of the caching strategy, i.e., the databases know that three out of five messages are chosen by the user, and for one of the chosen message, the first half of the message is cached, and for the remaining two chosen messages, the first $\frac{1}{4}$ portions are cached. However, the databases do not know which three messages are chosen.

In the retrieval phase, assume that the user wishes to retrieve message $W_1$ privately. For the first $\frac{1}{4}$ portion of messages, i.e., for the symbols in the interval $[0, \frac{L}{4}]$, since the user caches messages $W_2$, $W_3$ and $W_5$, the user applies the PIR scheme in \cite{chen2017capacity} with $M=3$. The total download cost for the first $\frac{1}{4}$  portion of the message, as shown in \eqref{eq_MDS_PIR}, is
\begin{align} \label{eq_ex3}
\frac{L}{4} \times \left(1+\frac{1}{2^{5-1-3}}\right).
\end{align}
For the following $\frac{1}{4}$ portion of messages, i.e., for the symbols in the interval $[\frac{L}{4}, \frac{L}{2}]$, since the user caches message $W_3$, the user applies the PIR scheme in \cite{chen2017capacity} with $M=1$. The total download cost for the second $\frac{1}{4}$  portion of the message, as shown in \eqref{eq_MDS_PIR}, is
\begin{align} \label{eq_ex5}
\frac{L}{4} \times \left(1+\frac{1}{2}+\frac{1}{2^2}+ \frac{1}{2^{5-1-1}}\right).
\end{align}
For the last half portion of messages, i.e., for the symbols in the interval $[\frac{L}{2}, L]$, since the user has not cached any messages, the user applies the PIR scheme in \cite{JafarPIR}. The total download cost for the last half portion of the message, as shown in \eqref{eq_PIR}, is
\begin{align} \label{eq_ex4}
\frac{L}{2} \times \left(1+\frac{1}{2}+\frac{1}{2^2}+\frac{1}{2^3}+\frac{1}{2^{5-1}} \right).
\end{align}

The overall download cost is the sum of \eqref{eq_ex3}, \eqref{eq_ex5} and \eqref{eq_ex4}. Therefore, the optimal normalized download cost is $\frac{29}{16}$, which can also be obtained through \eqref{eq_main2} by letting $r_1=\frac{1}{2}$, and $r_2=r_3=\frac{1}{4}$. Note that by applying the PIR scheme in \cite{chen2017capacity} to retrieve the first $\frac{1}{4}$ portion and the middle $\frac{1}{4}$ portion of the message, the databases cannot learn which messages have been cached by the user. In addition, both PIR schemes in \cite{JafarPIR} and \cite{chen2017capacity} hide the identity of the desired message. Therefore, the combination of these two PIR schemes is a feasible PIR scheme for PIR-PSI under a storage constraint \cite{arbmsgPIR}.

\subsection{General Achievable Scheme}

We now describe the general achievable scheme for $r_1 \geq r_2 \geq \dots \geq r_M$. We first consider the first $r_M$ fraction of messages, i.e., for the symbols in the interval $[0, Lr_M]$. Since $r_1 \geq r_2 \geq \dots \geq r_M$, the user caches $M$ messages for this portion. The user applies the PIR scheme in \cite{chen2017capacity} which results in the download cost
\begin{align} \label{eq_ach_1}
L r_M \times \left( 1+\frac{1}{N}++\frac{1}{N^2} +\dots  +\frac{1}{N^{K-1-M}}   \right).
\end{align}
Following the same logic, for the symbols in the interval $[Lr_i, Lr_{i-1}]$, $i\geq2$, the user caches $i$ messages for this portion. The user applies the PIR scheme in \cite{chen2017capacity} which results in the download cost
\begin{align} \label{eq_ach_2}
L (r_{i-1}-r_i) \times \left( 1+\frac{1}{N}++\frac{1}{N^2} +\dots  +\frac{1}{N^{K-i }}   \right).
\end{align}
Lastly, for the symbols in the interval $[Lr_1, L]$, the user caches no messages for this portion. The user applies the PIR scheme in \cite{JafarPIR} which results in the download cost
\begin{align} \label{eq_ach_3}
L (1-r_1) \times \left( 1+\frac{1}{N}++\frac{1}{N^2} +\dots  +\frac{1}{N^{K-1}}   \right).
\end{align}

The overall download cost is the sum of \eqref{eq_ach_1}, \eqref{eq_ach_2} for $i=2,3,\dots, M$, and \eqref{eq_ach_3}, which is \eqref{eq_main2}. By applying the PIR scheme in \cite{chen2017capacity} to retrieve symbols in the interval of $[0, Lr_1]$, the databases cannot learn which messages have been cached by the user. In addition, both PIR schemes in \cite{JafarPIR} and \cite{chen2017capacity} protect the identity of the desired message. Therefore, the combination of these two PIR schemes is a feasible PIR scheme for PIR-PSI under a storage constraint \cite{arbmsgPIR}.

\section{Conclusion}

In this paper, we have introduced a new PIR model, namely PIR-PSI under a storage constraint. In this model, the user randomly chooses $M$ messages and caches the first $r_i$ portion of the chosen messages for $i=1, \dots, M$ subject to the memory size constraint $\sum_{i=1}^M r_i=S$. In the retrieval phase, the user wishes to retrieve a message such that no individual database can learn the identity of the desired message and the identities of the cached messages. For each caching scheme, i.e., $(r_1, \dots, r_M)$, we characterized the optimal normalized download cost to be $D^*=1+\frac{1}{N}+\frac{1}{N^2}+\dots+\frac{1}{N^{K-1-M}}+\frac{1-r_M}{N^{K-M}}+\frac{1-r_{M-1}}{N^{K-M+1}}+\dots+\frac{1-r_1}{N^{K-1}}$. In addition, we showed that, for a fixed memory size $S$, and a fixed number of accessible messages $M$, uniform caching achieves the lowest normalized download cost, where uniform caching means $r_i=\frac{S}{M}$, $i=1,\dots, M$. Then, we showed that, for a fixed memory size $S$, among all $K-\left \lceil{S} \right \rceil+1$ uniform caching schemes, the uniform caching scheme caching $M=K$ messages achieves the lowest normalized download cost. Finally, we conclude that for a fixed memory size $S$, the uniform caching scheme caching $K$ messages achieves the lowest normalized download cost.

\end{document}